\documentstyle[12pt]{article}
\begin{document}
\title{A TEST FOR VARYING $G$}
\author{B.G. Sidharth\\
Centre for Applicable Mathematics \& Computer Sciences\\
Adarsh Nagar, Hyderabad - 500 063, India}
\date{}
\maketitle
\vspace{9 mm}
\begin{flushleft}
Key Words: Gravitational constant, Variation, Binary Pulsar
\end{flushleft}
\begin{flushleft}
PACS: 04.90.+e
\end{flushleft}
\begin{abstract}
In this note we consider a variable $G$ cosmology which is
consistent with observation and which had successfully predicted
an ever expanding accelerating universe. It is shown that the
observed shortening of the orbital periods of binary pulsars is
also in good agreement with this model.
\end{abstract}
In an earlier communication \cite{r1}, it was shown that a varying
$G$ cosmology
\begin{equation}
G \propto - \frac{G_0}{T},\label{e1}
\end{equation}
$T$ being the age of the universe, about 15 billion years,
correctly gives the precession of the perihelion of Mercury. In
this model particles are created out of the Quantum Vacuum in an
inflationary scenario type phase transition. This model has been
discussed in detail in the references cited \cite{r2,r3,r4,r5,r6}
and other references therein. In this cosmology at a given point
of time, within an uncertainty time interval $\tau$ of the order
of the Compton time of the typical elementary particle viz., the
pion given $N$ particles, $\sqrt{N}$ particles are created from
the Quantum Vacuum, whence
\begin{equation}
\sqrt{N}\tau = T\label{e2}
\end{equation}
Apart from other routine effects like the bending of light, the
cosmology correctly predicted an ever expanding accelerating
universe besides explaining many hitherto inexplicable features
like the so called large number coincidences, the mysterious pion
mass Hubble constant Weinberg relation and several other features
as discussed in the references.\\
We would now like to show that this model also explains the
observed decrease in the orbital period of the binary pulsar PSR
$1913 + 16$, otherwise
attributed to as yet undetected gravitational waves \cite{r7}.\\
In general in schemes in which $G$ the universal constant of
gravitation decreases with time, it is to be expected that
gradually the size of the orbit and the time period would
increase, with an overall decrease in energy. It may be mentioned
that cosmologies in which the constant of gravitation $G$ varies
with time have been considered in slightly different contexts also
\cite{r8,r9}. In any case all this becomes more relevant in the
light of latest observations that the fine structure constant
varies with time which also finds an explanation \cite{r10,r11}.\\
But in the present case as we will show, the gravitational energy
of the binary system, $\frac{GMm}{L}$ remains constant, where $M$
is the mass of the central object and $L$ the mean distance
between the objects. This is because the decrease in $G$ is
compensated by an increase in the material content of the system,
according to the
above model.\\
In fact the energy lost is given by $\frac{GM}{TL}$ (per unit mass
of the orbiting object - in any case the mass of the orbiting
object does not feature in the dynamical equations), on using
(\ref{e1}). Further as can be seen from (\ref{e2})
$\frac{1}{\sqrt{N}\tau} = \frac{1}{T}$ particles appear from the
Quantum Vacuum per second, per particle in the universe. So the
energy gained in this process is $\frac{GM}{TL}$ per second. This
follows, if we write $M = n \times m$, where $n$ is the number of
typical elementary particles in the central body and $m$ their
mass.\\
As can be seen from the above the energy lost per second is compensated by the energy
gained and thus the total gravitational energy of the binary system remains constant.\\
Let us now consider an object revolving about another object, as
in the case of the binary pulsar \cite{r12}. The gravitational
energy of the system is now given by,
$$\frac{GMm}{L} = const.$$
Whence
\begin{equation}
\frac{\mu}{L} \equiv \frac{GM}{L} = const.\label{e3}
\end{equation}
For variable $G$ we have
\begin{equation}
\mu = \mu_0 - tK\label{e4}
\end{equation}
where
\begin{equation}
K \equiv \dot \mu\label{e5}
\end{equation}
We take $\dot \mu$ to be a constant, in view of the fact that $G$
varies very slowly, as can be seen from (\ref{e1}).\\
To preserve (\ref{e3}), we should have
$$L = L_0 (1 - \alpha K)$$
Whence on using (\ref{e4})
\begin{equation}
\alpha = \frac{t}{\mu_0}\label{e6}
\end{equation}
We shall consider $t$, to be the period of revolution. Using
(\ref{e6}) it follows that
\begin{equation}
\delta L = - \frac{LtK}{\mu_0}\label{e7}
\end{equation}
We also know (Cf.ref.\cite{r12})
\begin{equation}
t = \frac{2 \pi}{h} L^2 = \frac{2\pi}{\sqrt{\mu}}\label{e8}
\end{equation}
\begin{equation}
t^2 = \frac{4\pi^2 L^3}{\mu}\label{e9}
\end{equation}
Using (\ref{e7}), (\ref{e8}) and (\ref{e9}), a little manipulation
gives
\begin{equation}
\delta t = - \frac{2t^2K}{\mu_0}\label{e10}
\end{equation}
(\ref{e7}) and (\ref{e10}) show that there is a decrease in the size of the orbit,
as also in the orbital period. Before proceeding further we note that such a
decrease in the orbital period has been observed in the case of binary
pulsars \cite{r7,r13}.\\
Let us now apply the above considerations to the case of the
binary pulsar PSR $1913 + 16$ observed by Taylor and co-workers
(Cf.ref.\cite{r13}). In this case it is known that, $t$ is 8 hours
while $v$, the orbital speed is $3 \times 10^7cms$ per second. It
is easy to calculate from the above
$$\mu_0 = 10^4 \times v^3 \sim 10^{26}$$
which gives $M \sim 10^{33}gms$, which of course agrees with
observation. Further we get using (\ref{e1}) and (\ref{e5})
\begin{equation}
\Delta t = \eta \times 10^{-5}sec/yr, \eta \leq 8\label{e11}
\end{equation}
Indeed (\ref{e11}) is in good agreement with the carefully observed
value of $\eta \approx 7.5$ (Cf.refs.\cite{r7,r13}).\\
Finally it may be remarked that this same effect has been
interpreted as being due to gravitational radiation, even though
there are some objections to the calculation in this case
(Cf.ref.\cite{r13}).

\end{document}